\documentclass[runningheads]{llncs}
\usepackage{tabularx}
\usepackage{graphicx}
\usepackage{subcaption}
\usepackage{wrapfig}
\usepackage{listings}
\begin{document}
\title{Boosted Random Forests for Predicting Treatment Failure of Chemotherapy Regimens}
\titlerunning{Boosted Forests for Chemotherapy Failure}
%
\author{Muhammad Usamah Shahid\inst{1}\orcidID{0009-0001-4293-2979}  \and
Muddassar Farooq\inst{1}}
\authorrunning{U. Shahid, M. Farooq}
%
\institute{CureMD Research, 80 Pine St 21st Floor, New York, NY 10005, United States
\email{\{muhammad.usamah,muddassar.farooq\}@curemd.com}\newline
\url{https://www.curemd.com/}}
\maketitle              
\begin{abstract}
Cancer patients may undergo lengthy and painful chemotherapy treatments, comprising several successive regimens or plans. Treatment inefficacy and other adverse events can lead to discontinuation (or failure) of these plans, or prematurely changing them, which results in a significant amount of physical, financial, and emotional toxicity to the patients and their families. In this work, we build treatment failure models based on the Real World Evidence (RWE) gathered from patients’ profiles available in our oncology EMR/EHR system. We also describe our feature engineering pipeline, experimental methods, and valuable insights obtained about treatment failures from trained models. We report our findings on five primary cancer types with the most frequent treatment failures (or discontinuations) to build unique and novel feature vectors from the clinical notes, diagnoses, and medications that are available in our oncology EMR. After following a novel three axes - performance, complexity and explainability - design exploration framework, boosted random forests are selected because they provide a baseline accuracy of 80\% and an F1 score of 75\%, with reduced model complexity, thus making them more interpretable to and usable by oncologists.

\end{abstract}
\section{Introduction}
Cancer patients undergo a lengthy and painful chemotherapy treatment that creates emotional toxicity, resulting in significant stress debt for the patients and their families. Unsuitable chemotherapy plans result in unpredictable and debilitating side effects that may lead to the death of patients in a worst case scenario. When oncologists detect an ineffective plan, it is discontinued and often replaced with a new one, even that adds financial toxicity to a patient (or their families) and the healthcare system. To enhance patient-centered quality of care, an assistive tool -- MedicalMind -- for oncologists is developed that can predict, at the time of chemotherapy selection, the likelihood of failure of a chemotherapy plan based on the real-world evidence (RWE) that is collected from the historical EMR data. It also generates clinically relevant explanations that will empower oncologists to take informed decisions to not only reduce the number of preventable deaths due to the side effects of chemotherapy but also save patients and their families from emotional and financial toxicity when chemotherapy plans are changed.

Most of the patients' information - MRI scans, pet scans, pathology slides and genomics data - used in clinical trials or focused studies, is not available in the EMR systems of oncology providers in the real world. The reason is that their major objective is to record clinical interventions - chemotherapy plans - and track their outcomes to gain reliable and valuable insights about their relative efficacy. Therefore, we are constrained to using the structured fields of EMR and providers' notes only which poses unique challenges -- noise, missing data fields, class imbalance -- including providers' habit of not providing useful information like tumor size, toxicity, and staging in the structured fields. 

Historically, mathematical models based on survival analysis are used to predict treatment failure by modeling tumor dynamics \cite{ribba2014review}. Their well-known shortcomings are broad generalizations that are an outcome of often simple and linear relations that are asserted by the models that are developed from clinical studies of small cohort sizes. Within the context of EHR data, text-based models including Bag-of-Words representation \cite{french2019identification} and Entropy-based categorization \cite{malakouti2019predicting} are also used. Sequential approaches that utilize states and events to find a correlation between two sequences \cite{silvina2019predicting} \cite{piernik2019assessing} are also used. Generally speaking, RNN based models outperform other sequential models, but their complexity and black-box nature make them unsuitable for use in medicine, as oncologists are clueless about the logic and reasoning  of a decision. 

Other direction of research focuses on predicting chemotherapy outcomes. The study presented in \cite{silvina2019predicting} uses toxicity as an outcome to predict chemotherapy treatment failure among breast cancer patients only. In \cite{cooper2020lung}, an instance-specific Bayesian model is reported for predicting survival in lung cancer patients. Tree ensemble methods have previously been used by researchers to predict treatment outcome in cancer patients. The authors of \cite{arezzo2021machine} accurately predict 24 month progression free survival in cervical cancer patients by using random forests; while the authors of \cite{kashef2020treatment} demonstrate the success of Xgboost in classifying treatment outcome of pediatric patients suffering from acute lymphoblastic leukemia. A study by \cite{moghadas2020priori} uses Adaboost to predict chemotherapy response for breast cancer patients by using biomarkers from CT images. 

In contrast, we work with simpler representations of EMR oncology data, where spatial snapshots of a patient's health and treatment profiles are taken at relevant time instances to look for hidden patterns. Therefore, after following a smart design exploration framework and extensive empirical evaluations, we discover that ensemble methods that combine boosting and bagging of decision trees by using boosted random forests \cite{mishina2015boosted} are the optimal choice for treatment failure use case in our multi-objective optimization landscape across three dimensions - performance, complexity and explainability. To the best of our knowledge, our treatment failure models are novel and unique for the oncology use case, as they not only are relatively accurate but also explain the reasons of their inference to oncologists. Our models provide a baseline accuracy of 80\% and an F1 score of 75\% for five most prevalent cancer types.

\section{Real World Evidence and Oncology Data Overview}
The medical records of cancer patients are collected from our Oncology EMR application. HIPAA guidelines are followed to anonymize and deidentify patient records to ensure the security and privacy of personally identifiable information. The registered patients have active treatments from 2015 to 2022. From a total of 21212 patients treated with chemotherapies, as shown in Table ~\ref{tab : data_overview}, we select the patients of top 5 cancer types that have the most frequent plan discontinuations.
\begin{table}
\vspace{-0.5cm}
	\caption{The total number of  patients and patients with discontinued treatments}
	\label{tab : data_overview}
	\centering
	\begin{tabular}{|c |p{0.4\textwidth}| c|c|}
		\hline
		ICD10 code & Description &  Patient Count & Treatment Failures \\
		\hline
		C18 & Malignant neoplasm of colon & 1034 & 332\\
		C34 & Malignant neoplasm of bronchia and lung &  1547 & 489\\
		C50 & Malignant neoplasm of breast &  2184 & 617\\
		C61 & Malignant neoplasm of prostate &  1074 & 361\\
		C90 & Plasmocytoma and malignant plasma cell neoplasms &  866 & 249\\
		\hline
	\end{tabular}
 \vspace{-0.5cm}
\end{table}
\begin{wraptable}{l}{0.6\textwidth}
\vspace{-0.5cm}
	\caption{Most frequent reasons for discontinuation of chemotherapy plans}
	\label{tab : disc_reasons}
	\centering
	\begin{tabular}{|p{0.4\textwidth} |c|}
		\hline
		
		Reason & \% failures \\
		\hline
		Change in therapy (drugs, doses or schedule) & 30.4\\
		Progression of disease &  18.0\\
		Adverse effects and other health concerns &  15.9\\
		No comment &  11.1\\
		Ambiguous (e.g. "MD order") &  10.3\\
		\hline
		Death or hospice & 2.0\\
		\hline
	\end{tabular}
 \vspace{-0.5cm}
\end{wraptable}

A chemotherapy plan comprises of multiple medications, dosages, and administration schedules. In our data, on average, patients are prescribed 3 plans over the course of their treatment, and approximately 40\% of the patients undergo only one chemotherapy plan. Overall 16-18\% of patients have at least one discontinued plan, and the most frequent reasons for discontinuations are tabulated in Table \ref{tab : disc_reasons}.  It is clear that 34\% of the discontinuations result due to poor treatment efficacy (disease progression or adverse effects) leading to the worst case outcome in 2\% of the cases: death of a patient or being referred to hospice. We now discuss our design exploration framework, beginning with feature engineering, that leads to an optimal treatment failure model in terms of its performance, complexity, and explainability.

\section{Feature Engineering} \label{sec: featureengineering}
We aim for a minimalist and relevant feature-set that captures dynamic changes in the diseases and treatment journeys of patients. Our feature engineering pipeline comprises of ingesting raw data of medical profiles of patients and then running autonomous transformation pipelines to convert it into processed information to enable feature engineering. Medical profiles of patients are grouped by cancer type which allows the creation of cancer-specific, unique features.

\textbf{Feature vector.}
For each patient, the data is ingested across multiple processed tables and arranged in a chronological (temporal) order. Chemotherapy plans are inferred from the orders by reviewing the medications present in them and by analyzing associated comments about therapy details and administration schedules. The designed features only consider a patient's profile to the date when the plan is being prescribed by oncologists. We only incorporate age and gender from a patient's demographics, as other information was either anonymized beforehand or scarcely available.

\begin{wraptable}{r}{5cm}
\vspace{-0.5cm}
	\caption{Distribution of normal (successful) and abnormal (discontinued) plans)}
	\label{tab : sample_count}
	\centering
	\begin{tabular}{|c|c|c|}
		\hline
		
		ICD10 Code & Normal & Abnormal \\
		\hline
		C18 & 1024 & 706\\
		C34 & 1466 & 945\\
		C50 & 2130 & 1345\\
		C61 & 914 & 685\\
		C90 & 796 & 593\\
		\hline
	\end{tabular}
 \vspace{-0.5cm}
\end{wraptable}
\textbf{Medications.}
The EMR tracks, at length, different drugs and their doses that are planned, ordered, administered (on-site infusion), and/or prescribed for use at home. Medications are identified by their name and a universal National Drug Codes (NDC) system. To reduce the dimensionality of our drug dataset, we use the Generic Product Identifier (GPI) \cite{kluwer2019medi} - a 14-character hierarchical classification system that identifies drugs from their primary therapeutic use down to the unique interchangeable product. Depending on relevance to a treatment, medications are abstracted to drug type (two digits), class (four digits) or sub-class (six digits). For a plan, each medication feature will have two occurrences:(1) total planned dosage if the proposed regimen were to be strictly followed; and (2) total dose administered to a patient in the last six months to analyze the impact of a treatment or its side effects.

\textbf{Diagnoses.}
 International Classification of Diseases (ICD10) codes are used and collapsed into thirty different Elixhauser comorbidity groups \cite{elixhauser1998comorbidity}. Three types of features are then extracted from a list of the patient's active diagnoses at the time of selecting a plan: (1) certain ICD codes e.g. R97.2, which encodes elevated levels of prostate-specific antigen, are used directly as features by searching whether they are present or not in the patients' records; (2) important comorbidity conditions e.g. renal failure are marked present or absent from the Elixhauser comorbidity groups derived from the ICD10 codes; and (3) Elixhauser readmission score, computed at the time of selecting a chemotherapy plan, and how it has changed in the last six months.

\textbf{Vitals and Lab results.}
At the time of selecting a plan, the last reading of a desired vital or lab test is used in the feature vector. Exponentially weighted moving averages are used for imputation and noise removal, complemented by substituting population means or clinically normal values for the patients with no relevant information stored in EMR. Currently, only body surface area and serum creatinine values are used as features, as they are recorded for approximately all of the patients in the EMR system.

\textbf{Notes.}
Critical information like staging, tumor information (size, grade, risk), ECOG or Karnofsky performance values, and biomarkers that are needed to select the type of chemotherapy are either completely missing or scarcely provided in the structured fields by oncologists. Luckily, they report most of these features in the clinical notes. We built a novel smart annotation engine that automatically annotates the oncologists' notes and subsequently extracts desired features from the annotated text. The detailed description of feature extraction from unstructured oncology notes is beyond the scope of this paper.

We curate cancer-specific and relevant feature-set by identifying and filtering redundant or non-informative features before the sampling stage.

\textbf{Sampling.}
Plans are sampled for each patient before computing feature vectors or splitting the data for training and evaluation. The strategy is chosen to minimize class imbalance and remove bias in features. After ensuring the plan is prescribed for a chosen diagnosis, all unique chemotherapy plans, excluding the first one, are sampled for patients with no discontinuations. Abnormal samples are obtained by sampling discontinued plans and pairing them with the last unique plan prior to the discontinuation, if any. The number of samples in the two classes - abnormal and normal - is reported in Table \ref{tab : sample_count}.

\section{An Empirical Design Exploration Framework}
Our design exploration framework consists of three steps:(1) select well known machine learning classifiers from each learning paradigm and shortlist top 3 models that perform the best on our oncology dataset; (2) deep dive into the shortlisted models by exploring the best hyperparameter settings and select the best model; (3) build model insights for explainability to oncologists. 
\begin{table}[t]
\vspace{-0.5cm}
	\caption{F1 scores (\%) on the validation dataset, with top 3 highlighted}
	\label{tab : val_res_other}
	\centering
\begin{tabular}{|c|c|c|c|c|c|c|c|c|c|c|c|}
		\hline
		Cancer & Log & Elastic & KNN & SVM & SVM & Tree   & Naive  & Ada- & Random & Xgb- & Boosted\\
		type & reg & net & (k=5) & linear & rbf & (D=10)  & Bayes  & boost & forest & oost & forest\\
		\hline
		C18 & 54.2  & 58.3  & 52.6   & 52.4  & 59.6  & 65.7   & 59.6   & 64.3 & \textbf{75.3} & \textbf{69.6} & \textbf{76.4}\\
		C34 & 55.3  & 60.9  & 52.0   & 50.0  & 55.4  & 64.6   & 54.3   & 62.5 & \textbf{76.3} & \textbf{76.2} & \textbf{77.5} \\
		C50 & 59.0  & 60.5  & 61.5   & 52.8  & 59.9  & 69.0   & 19.6   & 66.0 & \textbf{81.9} & \textbf{79.0} & \textbf{82.4}\\
		C61 & 49.7  & 57.1  & 57.7   & 15.5  & 51.1  & 60.8  & 61.8   & 63.1 & \textbf{72.0} & \textbf{72.4} & \textbf{72.9} \\
		C90 & 65.5  & 60.6  & 59.1   & 49.3  & 62.1  & 77.6  & 34.5   & 77.4 & \textbf{84.6} & \textbf{80.9} & \textbf{85.9}\\
		\hline
		
	\end{tabular}
 \vspace{-0.1cm}
\end{table}

\textbf{Step 1: Shortlist top 3 Models}. We select the following well-known machine learning classifiers to begin with for doing a comprehensive performance evaluation:(1) logistic regression, first with no penalties and then with combined l1 and l2 penalties (Elastic Net); (2) k Nearest neighbors (KNN); (3) SVM with linear and RBF kernels; (4) single decision tree, with a depth cut-off of 10; (5) Naive Bayes classifier; (6) Adaboost classifier with 100 decision stumps; and (7) the three tree ensemble methods -- random forest, Xgboost, boosted forest -- with 500 trees each.

Separate models are trained for each cancer type, with a three-way data split: 75\% for training, 15\% for validation and 10\% for testing. The splits are random, but fixed, for each model over multiple iterations of training; and class distributions are also kept constant in all splits. To account for the randomness of some of the models, multiple instances of each model are trained and the model that performs the best on the validation set is selected. As is evident from Table \ref{tab : sample_count}, a significant imbalance exists between two classes; therefore, F1 score is more appropriate to benchmark the performance of different classifiers. The results are tabulated in Table \ref{tab : val_res_other}. It is obvious that random forests (RF), Xgboost (XGB) and boosted forests (BF) - all ensembles of trees - are the best performing models for our treatment failure application.

\textbf{Step 2: Hyperparameter Settings to select the best Model.}
\begin{wraptable}{r}{4.5cm}
\vspace{-0.5cm}
	\caption{Hyperparameter settings for tree ensemble experiments}
	\label{tab : bf_hyper}
	\centering
	\begin{tabular}{|c|c|c|c|c|}
		\hline
		
		Identifier & $N$ & $n_f$ & $n_t$ & $D$ \\
		\hline
		HP1 & 500 & 10 & 50 & None\\
		HP2 & 100 & 10 & 10 & None\\
		HP3 & 100 & 10 & 10 & 10\\
		HP4 & 50 & 5 & 10 & 10\\
		HP5 & 25 & 5 & 5 & 10\\
		HP6 & 10 & 2 & 5 & 10\\
		\hline
	\end{tabular}
 \vspace{-0.5cm}
\end{wraptable}
We provide a description of the hyperparameters that are used for all three ensembles on the validation set, and how the respective boosted forest is constructed in Table \ref{tab : bf_hyper}, where $N$ represents the number of trees and $D$ the depth cut-off of each tree in each ensemble model. For boosted forests, $n_f$ and $n_t$ are the number of forests and trees in each forest respectively. Table \ref{tab : val_res_trees} tabulates the F1 scores of three ensemble methods on the validation set. Please note that the best results are marked in bold. It is obvious that a designer needs to trade performance for complexity and interpretability i.e. as he selects relatively less complex models that have better explainability (HP6 over HP1), the performance degrades from 3-7\%. Generally speaking, the results validate the hypothesis that boosted forests outperform other ensemble methods for the majority of hyperparameter settings for all five cancer types on our oncology dataset.

\begin{wrapfigure}{r}{5cm}
\vspace{-0.5cm}
\centering
	\includegraphics[width=5cm, clip]{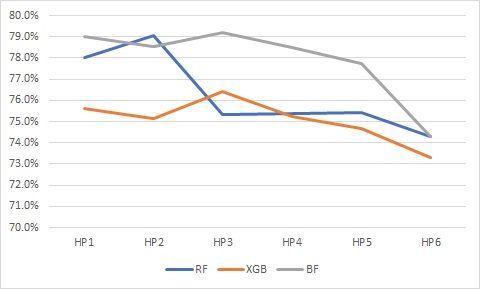}
	\caption{Average F1 scores of 3 ensemble methods} \label{fig : ave_f1}
 \vspace{-0.5cm}
\end{wrapfigure}

To better visualize the impact of hyperparameter settings on three ensemble methods, we analyze the average F1 scores for each hyperparameter setting on the validation sets and plot the average in Figure \ref{fig : ave_f1}. It is interesting to note that once the tree depth is capped, the performance of random Forests is significantly degraded, whereas the performance of boosted ensembles slightly increase. Boosted forests provide superior performance compared with the other two except for HP2 configuration.  We also notice a common "elbow" on HP5, suggesting that these hyperparameter settings ($N = 25, D = 10$) are the best compromise between model interpretability and performance.

\begin{table}[t]
\vspace{-0.5cm}
	\caption{F1 scores (\%) for 3 ensemble methods on the validation set}
	\label{tab : val_res_trees}
	\centering
 \resizebox{\columnwidth}{!}{%
	\begin{tabular}{|c|ccc|ccc|ccc|ccc|ccc|ccc|}
		\hline
		Cancer & & HP1 & & & HP2 & & & HP3 & &  & HP4 & &  & HP5 & &  & HP6 & \\
		type & RF    & XGB   & BF    & RF    & XGB   & BF    & RF    & XGB   & BF    & RF    & XGB   & BF    & RF    & XGB   & BF    & RF    & XGB & BF \\
		\hline
		C18 & 75.3  & 69.6  & \textbf{76.4}  & \textbf{76.3}  & 69.3  & 74.4  & 75.0  & 73.4  & \textbf{75.4}  & \textbf{75.3}  & 73.1  & 75.1  & 74.2  & 72.9  & \textbf{75.5}  & 72.9  & \textbf{73.5}  & 73.2 \\
		C34 & 76.3  & 76.2  & \textbf{77.5}  & \textbf{77.6}  & 75.5  & 76.7  & 72.1  & 77.0  & \textbf{77.9}  & 73.4  & 75.4  & \textbf{76.8}  & 72.7  & 74.6  & \textbf{75.2}  & 69.7  & \textbf{71.4}  & 70.0 \\
		C50 & 81.9  & 79.0  & \textbf{82.4}  & 81.8  & 80.8  & \textbf{82.1}  & 78.0  & 79.0  & \textbf{82.0 } & 77.2  & 78.7  & \textbf{81.1}  & 77.9  & 77.3  & \textbf{80.1}  & 77.2  & 75.8  & \textbf{75.9} \\
		C61 & 72.0  & 72.4  & \textbf{72.9}  & 73.8  & 68.4  & \textbf{73.9}  & 70.8  & 70.1  & \textbf{73.8}  & 70.0  & 68.4  & \textbf{73.4}  & 70.3  & 69.7  & \textbf{72.8}  & 69.8  & 68.7  & \textbf{71.4} \\
		C90 & 84.6 & 80.9 & \textbf{85.9} & \textbf{85.9} & 81.6 & 85.5 & 80.7 & 82.6 & \textbf{87.0} & 81.0 & 80.7 & \textbf{86.0} & 82.1 & 78.7 & \textbf{85.1} & \textbf{81.8} & 77.2 & 80.9 \\
		\hline
	\end{tabular}
 }
 \vspace{-0.1cm}
\end{table}

\begin{wrapfigure}{r}{5.5cm}
\vspace{-0.5cm}
\centering
	\includegraphics[width=5.5cm, clip]{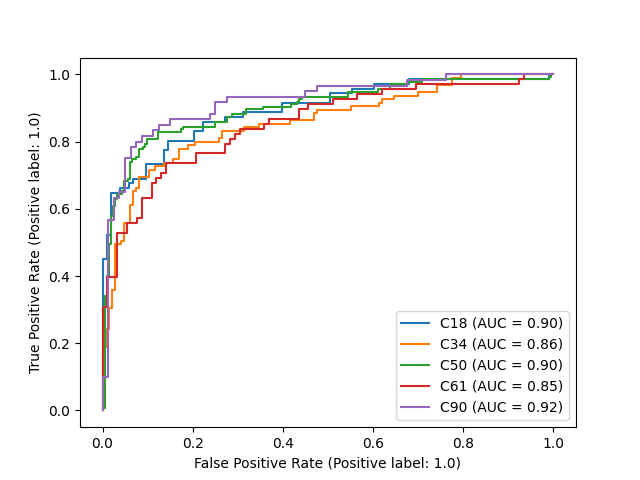}
	\caption{ROC curves for the boosted forests calculated on the test dataset } \label{fig : roc}
 \vspace{-0.5cm}
\end{wrapfigure}

From Step 2, we select the boosted forests models and evaluate all of their performance metrics - accuracy, precision, recall, specificity and F1 score - on the test dataset and tabulate the results in Table \ref{tab : bf_test} on all five cancer datasets. It is safer to claim that the selected models maintain a baseline accuracy of 80\% and F1 score of 75\% to predict treatment failure of chemotherapy plans for five cancer types. Our models generally have much better specificity and precision than recall and this is a general symptom of imbalanced classes. But it also means that false positives are highly unlikely to occur although false negatives may happen, especially in our lung cancer model. The ROC curves in Figure \ref{fig : roc} suggest that explored models are very good at distinguishing between successful and failed chemotherapy plans. All five cancer models have an AUC value of 0.85 or more.

\textbf{Step 3: Model insights and explainability.}
The boosted forest models with the HP5 hyperparameter setting are the optimal models that are selected from Step 2 of our design exploration framework. We now provide model insights and discuss explainability.

\begin{table}[t]
\vspace{-0.5cm}
	\caption{Results (in \%) of the chosen boosted forest model on the test set}
	\label{tab : bf_test}
	\centering
	\begin{tabular}{|c|c|c|c|c|c|c|c|}
		\hline
		
		Cancer type & Accuracy & Precision & Recall & NPV & Specificity & F1 score & AUROC\\
		\hline
		C18 & 83.3 &	83.9 &	73.2 & 83.0 & 90.3 & 78.2 & 89.5\\
		C34 & 81.0 &	85.5 &	62.1 & 79.2 & 93.2 & 72.0 & 85.8\\
		C50 & 86.5 &	86.1 &	77.8 & 86.7 & 92.1 & 81.7 & 90.0\\
		C61 & 80.6 &	79.4 &	73.6 & 81.4 & 85.9 & 76.3 & 85.5\\
		C90 & 84.8 &	84.5 &	81.7 & 86.6 & 88.8 & 83.1 & 91.9\\
		\hline
	\end{tabular}
 \vspace{-0.1cm}
\end{table}

\begin{figure}
	\centering
	\begin{subfigure}[b]{0.35\textwidth}
		\includegraphics[width=\textwidth]{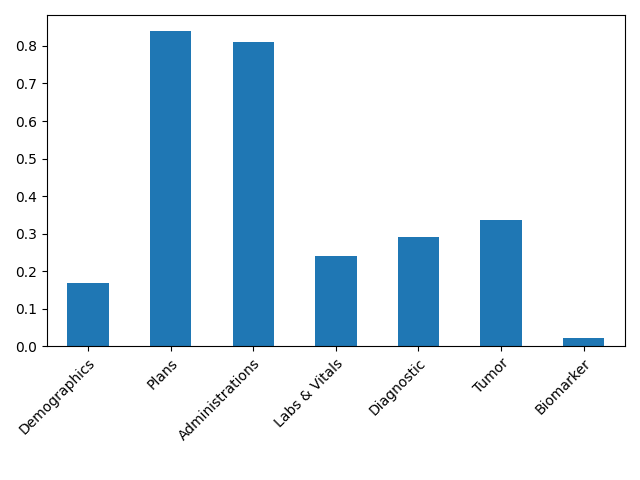}
		\caption{Sum of MAD scores - how important is each category} \label{fig : sum_mad}
	\end{subfigure}
\begin{subfigure}[b]{0.35\textwidth}
	\includegraphics[width=\textwidth]{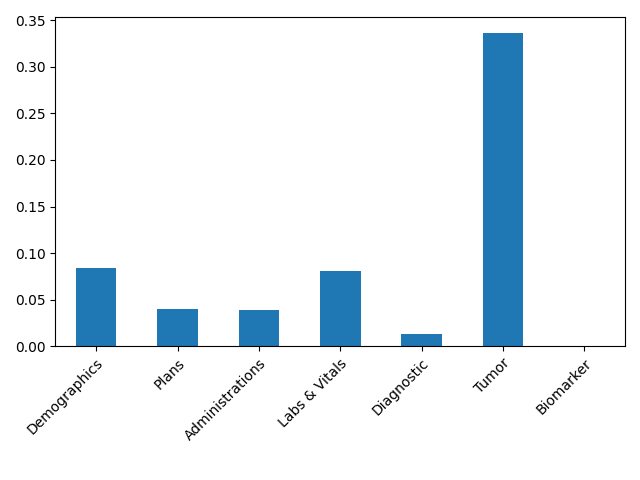}
	\caption{Average MAD scores of each feature in the category} \label{fig : mean_mad}
\end{subfigure}
	\caption{Feature importance by categories in the selected boosted forest model}
\end{figure}

\textbf{Feature importance.}
Though we have engineered unique feature sets for each cancer type, still we can discuss important features by grouping them into broad categories as discussed in Section \ref{sec: featureengineering}. We permute feature importance to calculate the mean accuracy decrease (MAD), if the value of a feature is randomly toggled. To compare the categories, we present the sum and the average of the MAD scores of all features belonging to each category, across all five models in Figures \ref{fig : sum_mad} and \ref{fig : mean_mad}. On the average, tumor features -- tumor size, staging, grade and risk -- are among the most important features. Cumulatively, dosages of planned medications and past administrations, at the time of selecting a plan, have had the greatest impact on the predictions of our boosted forest models. Biomarkers - information extracted from notes about active mutations - make the least contributions to treatment failure decisions, perhaps due to a scarce availability even in the notes.

\textbf{Explanations of Treatment Failures Models.}
One logical benefit of using boosted forests is that we can easily combine concepts of classifier weights, probabilistic output of trees and forward chaining to generate less number of simpler, yet powerful rules to explain the inference of our models. We use the following 3 step algorithm to shortlist rules:
\begin{figure}

	\centering
	\begin{subfigure}[b]{0.35\textwidth}
		\includegraphics[width=\textwidth]{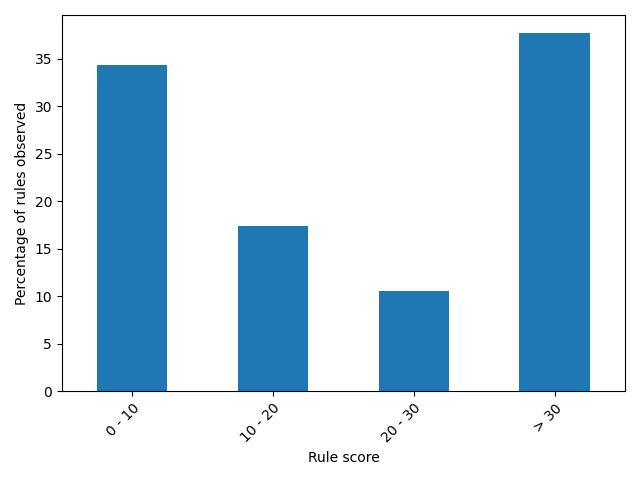}
		
	\end{subfigure}
	\begin{subfigure}[b]{0.35\textwidth}
		\includegraphics[width=\textwidth]{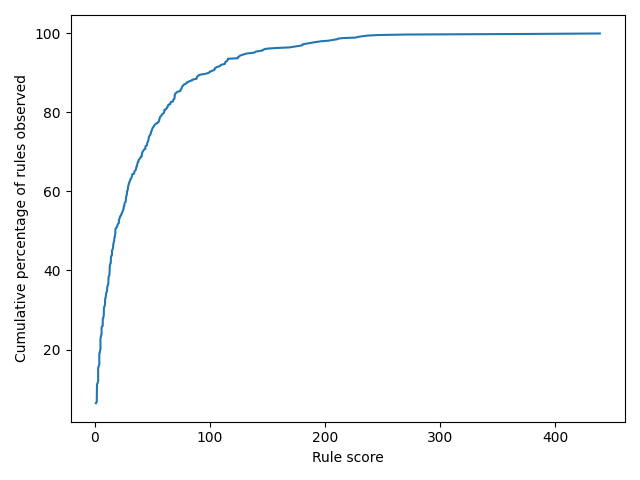}
		
	\end{subfigure}
	\caption{Trends observed in rule scores of the 50 test samples}\label{fig : rule_scores}
\end{figure}
\begin{enumerate}
	\item Drop forests whose predictions contradict with the prediction of the boosted ensemble model;
	\item In the remaining forests, drop trees that predict an opposite outcome than the final prediction made by the model;
	\item Rules, extracted by forward chaining on the remaining trees, are sorted by their importance or strength. We define a simple "rule score" as the product of the class probability determined by the leaf nodes of a tree and the residual number of samples at that leaf node.
\end{enumerate}
Using this rule pruning algorithm, oncologists are presented with an ordered set of the most relevant and strong rules, for our treatment failure use case.

We examine 10 random test samples from each cancer type. On an average, 4 out of 5 forests and 15 of the 25 individual trees are in agreement with the final prediction of the models. If we analyze the rule scores, only 6 per predictions have had a score above 30 (see Figure \ref{fig : rule_scores}). The trends observed in the rules are summarized in Figure \ref{fig : rule_scores}. The threshold to shortlist rules based on their importance or strength, computed from the real world evidence (RWE) in the data, may be arbitrarily chosen to meet the requirements of oncologists. The higher the value of "rule score" is, the stronger the real world evidence that it represents.

Listing \ref{lst} shows 3 handpicked rules from our models of varying strengths. A simple text matching and replacement strategy can translate feature names and thresholds to a natural language description of a rule that oncologists are accustomed to reading as clinical guidelines. 

\begin{lstlisting}[caption={Example rules from the boosted forest model},captionpos=b,frame=lrtb,label={lst}]
If: tumor size <=4.3 cm and 
    Elixhauser readmission score < 39 and the plan 
    comprises of <= 39.8 g of antimetabolites, 
    <= 0.22 g of Opdivo, <= 0.07 g of corticosteroids and 
    <= 1.35 g of antihistamines, 
    and in last six months patient has received
    <= 4.79 g of antimetabolites, <= 0.001 g of Topoiso-
    merase I inhibitors and <= 0.021 g corticosteroids
Then Prediction: Treatment will fail 
     Rule Score = 140

 If: in the last six months patient has received
    <= 1.8 g of antimetabolites 
    and has not been diagnosed for adverse effect or
    underdosing of antineoplastic/immunosuppressive drugs
Then Prediction: Treatment will succeed 
     Rule Score = 69

If: the patient is less than 36 years old 
    and in the last six months patient has received 
    > 1.8 g of antimetabolites
Then Prediction: Treatment will fail
     Rule Score: 11

\end{lstlisting}
\section{Conclusion}
In this paper, we follow a customized design exploration framework to select optimal models for a novel use case of predicting failure of chemotherapy regimens by using only the data that is available in an outpatient oncology EMR system. Our feature engineering pipeline is discussed to obtain distinctive feature sets for five most prevalent primary cancers. Our studies demonstrate that boosted forests, which are little investigated in the prior art, provide the optimal models in our multi-objective exploration space of performance, complexity and interpretability. The selected models provide a baseline accuracy of 85\%, F1 score of 75\%, and AUC of 0.85 on the test data. Finally, we explain models by discussing feature importance and the strength of shortlisted rules.

In the future, we want to validate the system in controlled clinical trials. The proposed method can also be adapted to treatments of various diseases, accounting for the need of engineering customized features, a time-consuming task. These hand-crafted features require understanding of domain knowledge, and consequently cannot be directly applied for other applications. However, a human-in-the-loop is necessary for ascertaining clinical relevance for AI in medicine tasks.

\bibliographystyle{splncs04}
\bibliography{bibliography}

\end{document}